\documentclass[letterpaper, 10 pt, conference]{ieeeconf}  % Comment this line out if you need a4paper

\IEEEoverridecommandlockouts                              % This command is only needed if 
\usepackage{graphicx}      % include this line if your document contains figures

\usepackage{subfigure}
\usepackage{float}
\usepackage[ruled]{algorithm2e}
\usepackage{amssymb}     
\usepackage{amsmath}
\usepackage{bm}
\usepackage{booktabs} 
\usepackage{cite} 
\usepackage{color} 

\newcommand{\pushright}[1]{\ifmeasuring@#1\else\omit\hfill$\displaystyle#1$\fi\ignorespaces}

\newtheorem{theorem}{Theorem}[section]

\newtheorem{definition}{Definition}[section]

\newtheorem{assumption}{Assumption}[section]

\title{\LARGE \bf
	Event-triggered Approximate Byzantine Consensus \\with Multi-hop Communication
}

\author{Liwei~Yuan and~Hideaki~Ishii 
	\thanks{L. Yuan and H. Ishii are with the Department of Computer Science, Tokyo Institute of Technology, Yokohama, 226-8502, Japan. e-mail: {\tt\small yuan@sc.dis.titech.ac.jp}, {\tt\small ishii@c.titech.ac.jp}.\newline
		\hspace*{1.8mm} This work was supported in the part by JSPS under Grant-in-Aid for 
		Scientific Research Grant No.~18H01460. The support provided by the China
		Scholarship Council is also acknowledged.
	}%
}

\begin{document}

	\maketitle
	\thispagestyle{empty}
	\pagestyle{empty}

	%%%%%%%%%%%%%%%%%%%%%%%%%%%%%%%%%%%%%%%%%%%%%%%%%%%%%%%%%%%%%%%%%%%%%%%%%%%%%%%%
	\begin{abstract}
		
		In this paper, we consider a resilient consensus problem
		for the multi-agent network where some of the agents are subject to Byzantine attacks and may transmit erroneous state values to their neighbors. 
		In particular, we develop an event-triggered update rule to tackle this problem as well as reduce the communication for each agent.
		Our approach is based on the mean subsequence reduced (MSR) algorithm with agents being capable to communicate with multi-hop neighbors.
		Since delays are critical in such an environment, 
		we provide necessary graph conditions for the proposed algorithm to perform well with delays in the communication.
		We highlight that through multi-hop communication,
		the network connectivity can be reduced especially in comparison
		with the common one-hop communication case.
		Lastly, we show the effectiveness of the proposed algorithm by a numerical example.
		
	\end{abstract}

	\section{Introduction}

	As concerns for cyber security have rised in general, multi-agent consensus problems in the presence of adversary agents creating failures and attacks have attracted much attention; see, e.g.,
	\cite{leblanc2013resilient, su2017reaching, nugraha2020dynamic, kikuya2018fault}. 
	One class of interdisciplinary problems that have been studied in both control and computer science is that of resilient consensus \cite{dibaji2017resilient, leblanc2013resilient, vaidya2012iterative}. 
	In these works, the adversary agents are categorized into basically two types: Malicious agents and Byzantine agents. These agents are capable to manipulate their data 
	arbitrarily. Malicious agents are limited as they must broadcast the same messages 
	to their neighbors, while Byzantine agents are capable to send individual messages to different neighbors (e.g., \cite{leblanc2013resilient, Lynch}).

	In this paper, we study the approximate Byzantine consensus using a mean subsequence reduced (MSR) algorithm. Such algorithms have been well studied in the fields of fault-tolerant techniques for multi-agent systems (e.g., \cite{azadmanesh2002asynchronous, leblanc2013resilient, vaidya2012iterative}). A basic assumption in MSR algorithms is the knowledge regarding an upper bound on the maximum number of malicious agents among the neighbors; this bound is denoted by $f$ throughout this paper. Then, at each iteration, each node removes the $f$ largest values and $f$ smallest values from neighbors to avoid being influenced by such potentially faulty values.
	Moreover, the graph property called robustness is shown to be critical for the network structure, guaranteeing the success of resilient consensus algorithms \cite{dibaji2017resilient, leblanc2013resilient}. In \cite{vaidya2012iterative}, the authors proposed a tight necessary and sufficient condition for Byzantine consensus, where such a condition can also be interpreted using the notion of robustness.
	However, such robustness requires the network to be relatively dense and complex. Therefore, how to enhance resilience of a sparse network without changing the original network topology has become an urgent problem.
	
	There are several works that tackled this problem by introducing the multi-hop communication techniques \cite{su2017reaching}, \cite{sakavalas2020asynchronous}, \cite{yuan2021resilient}. 
	Multi-hop communication techniques are commonly used in the areas of wireless communication \cite{goldsmith2005wireless}, computer science \cite{Lynch}, and systems control \cite{zhao2016global}. It is clear that with multi-hop communication, each node can have more information for updates compared to the one-hop case. Thus, the network may have more resilience against adversary nodes. For instance, 
	the works \cite{yuan2021secure, zhao2018resilient} pursued an approach based on detection of malicious agents in the network. Compared to MSR algorithms, which do not have such detection capabilities, the algorithms are applicable to more sparse networks with the same tolerance against malicious agents.
	Furthermore, in \cite{su2017reaching}, by introducing multi-hop communication in MSR algorithms, the authors solved the Byzantine consensus problem with a weaker condition on network structures compared to that derived under the one-hop communication model \cite{vaidya2012iterative}. 
	In \cite{sakavalas2020asynchronous}, the authors studied the asynchronous Byzantine consensus based on a flooding algorithm, where nodes relay their values over the entire network. Moreover, in our previous work \cite{yuan2022asynchronous}, we studied the asynchronous Byzantine consensus using an algorithm which is of less complexity than that in \cite{sakavalas2020asynchronous}. To conclude, through multi-hop communication, the connectivity requirement becomes less stringent for guaranteeing the same level of resilience as for the one-hop case. This is enabled by increasing the amount of data exchanged among agents through message relaying.

	In this paper, we aim to reduce the transmissions for the agents using the multi-hop weighted MSR algorithm \cite{yuan2021resilient} through event-triggered protocols \cite{heemels2012introduction}. Event-based protocols have been developed for
	conventional consensus without adversary agents in, e.g., \cite{dimarogonas2012distributed,kadowaki2015event,mishra2022event}. Moreover, the work \cite{wang2019resilient} proposed two event-based MSR algorithms using one-hop communication to reduce the transmissions. Among these works, event-triggered schemes have shown their effectiveness in reducing the transmissions for the agents using distributed algorithms even under adversarial environments. Moreover, time delays can be a critical factor affecting the performance of agents in the multi-hop communication. Hence, we introduce event-triggered protocols to the multi-hop weighted MSR algorithm, and we are interested to analyze the performance of the proposed algorithm with delays in the communication between agents. 
	Agents using the event-triggered multi-hop MSR algorithm will update locally, and they send their own state values along with relayed values only when the difference between the current value and the past communicated value exceeds a given threshold. Through simulations, we can see that the agents' transmissions can be significantly reduced compared to the multi-hop algorithm without the event-triggered protocol \cite{yuan2022asynchronous}. Furthermore, compared to the one-hop MSR algorithm with or without event-triggered protocols \cite{wang2019resilient}, \cite{leblanc2013resilient}, the connectivity requirement for our algorithm is less stringent. Besides, we analyze the performance of our algorithm with delays in communication, which is a case not studied in \cite{wang2019resilient}.

	The rest of this paper is organized as follows. 
	Section~II outlines preliminaries on graphs and the system model. Section~III presents the event-triggered multi-hop MSR algorithm and the definition of strongly robust graphs with multi-hop communication.
	In Section IV, we derive a condition under which the proposed algorithm reaches resilient consensus under asynchronous updates with delays.
	Section~V provides numerical examples to show the effectiveness of the proposed algorithm.
	Lastly, Section~VI concludes the paper.

	\section{Preliminaries}

	\subsection{Network Model}
	Consider the directed graph $\mathcal{G} = (\mathcal{V},\mathcal{E})$ consisting of the node set $\mathcal{V}=\{1,...,n\}$ and the edge set $\mathcal{E}\subset \mathcal{V} \times \mathcal{V}$. The edge $(j,i)\in \mathcal{E}$ indicates that node $i$ can get information from node $j$. 
	%A complete graph denoted by $\mathcal{K}_n= (\mathcal{V},\mathcal{E})$ is defined by $\mathcal{E} = \{(j, i)\in \mathcal{V} \times \mathcal{V} : i \neq j\}$.
	A path from node $i_1$ to $i_m$ is a sequence of distinct nodes $(i_1, i_2, \dots, i_m)$, where $(i_j, i_{j+1})\in \mathcal{E} $ for $j=1, \dots, m-1$. Such a path is referred to as an $(m-1)$-hop path (or a path of length $m-1$) and also as $(i_1,i_m)$-path when length is not relevant but the source and destination nodes are. We also say that node $i_m$ is reachable from node $i_1$. 
	%It may differ from effective topology of $\mathcal{G}$ in an adversarial environment. \footnote[2]{\textit{Effective topology} refers to the network pattern where edges represent the availability of the correct information of the connected agents. See Section \ref{Robustness} for more details.} 
	
	For node $i$, let $\mathcal{N}_i^{l-}$ be the set of nodes that can reach node $i$ via at most $l$-hop paths, where $l$ is a positive integer. Also, let $\mathcal{N}_i^{l+}$ be the set of nodes that are reachable from node $i$ via at most $l$-hop paths. 
	%Note that we let $i \in \mathcal{N}_i^{l-}$ and  $i \in \mathcal{N}_i^{l+}$. 
	%When $l=1$, we write $\mathcal{N}_i^{1-}$ and $\mathcal{N}_i^{1+}$ as $\mathcal{N}_i^{-}$ and $\mathcal{N}_i^{+}$, which are the sets of the direct in-neighbors and out-neighbors of node $i$.
	The $l$-th power of the graph $\mathcal{G}$, denoted by $\mathcal{G}^l$, is a multigraph\footnote[1]{
		In a multigraph, two nodes can have multiple edges between them.} with the same vertices as $\mathcal{G}$ and a directed edge from node $j$ to node $i$ is defined by a path of length at most $l$ from $j$ to $i$ in $\mathcal{G}$. 
	The adjacency matrix $A = [a_{ij} ]$ of $\mathcal{G}^l$ is given by $\alpha \leq a_{ij}<1$ if $j\in \mathcal{N}_i^{l-}$ and otherwise $a_{ij} = 0$, where $\alpha > 0$ is a fixed lower bound. We assume that $\sum_{j=1,j\neq i}^{n} a_{ij}\leq 1$. Let $L = [b_{ij} ]$ be the Laplacian matrix of $\mathcal{G}^l$, whose entries are defined as $b_{ii} =\sum_{j=1,j\neq i}^{n}a_{ij}$ and $b_{ij} = -a_{ij}$ for $ i\neq j$; we can see that the sum of the elements of each row of $L$ is zero.

	Node $i_1$ can send messages of its own to its $l$-hop neighbor $i_{l+1}$ via different paths.
	We represent a message as a tuple $m=(w,P)$, where $w=\mathrm{value}(m)\in \mathbb{R}$ is the message content, and $P=\mathrm{path}(m)$ indicates the path via which message $m$ is transmitted. 
	Moreover, nodes $i_1$ and $i_{l+1}$ are the message source and destination, respectively.
	When the source $i_1$ sends the message, $P$ is a path vector of length $l+1$ with the source being $i_1$ and other entries being empty. Then the one-hop neighbor $i_2$ receives this message from $i_1$, and it stores the value of node $i_1$ for consensus and relays the value of node $i_1$ to all the one-hop neighbors of $i_2$ with the second entry of $P$ being $i_2$ and other entries unchanged. This relay procedure will continue until every entry of $P$ of this message is occupied, i.e., this message reaches node $i_{l+1}$. We denote by $\mathcal{V}(P)$ the set of nodes in $P$.

\subsection{Update Rule}\label{problemsetting}

In graph $\mathcal{G} = (\mathcal{V},\mathcal{E})$,
the node set $\mathcal{V}$ is partitioned into the set of normal nodes $\mathcal{N}$ and the set of adversary nodes $\mathcal{A}$, where $|\mathcal{N}|=N$. The partition is unknown to the normal nodes at all times.

The update rule for normal agent $i$ is described by
\begin{equation}
x_i[k + 1] = x_i[k] + u_i[k],  \label{system_syn}
\end{equation}
where $x_i[k] \in \mathbb{R}$ is the state and $u_i[k]$ is the control input given by
\begin{equation}
u_i[k]=\sum_{j\in \mathcal{N}_i[k]} a_{ij}[k](\hat{x}_j[k]-x_i[k]). \label{input_syn}
\end{equation}
Here, $\hat{x}_j[k] \in \mathbb{R}$ is an auxiliary state, representing the last communicated state of node $j$ at time $k$. It is defined as
\begin{equation}
\hat{x}_j[k] = x_j[t_h^j], k\in [t_h^j,t_{h+1}^j),
\end{equation}
where $t_0^j, t_1^j, \dots $ denote the transmission times of node $j$ determined by the triggering function to be given below. The initial values $x_i[0], x_j [0]$ are given, and $a_{ij}[k]$ is the weight for the edge $(j, i)$. Note that at initial time, $\hat{x}_i[0]$ need not be the same as $x_i[0]$.
Let $a_{ii}[k] = 1 - \sum_{j\in \mathcal{N}_i^{l-}[k]} a_{ij}[k]$. Assume that $\gamma\leq a_{ij}[k]<1$ if $a_{ij}[k]\neq 0$ or $i = j$ for $i, j \in \mathcal{V}$, where $0<\gamma<1$. In the resilient consensus algorithm to be introduced, the neighbors whose values are used for updates change over time and, hence, the weights $a_{ij}[k]$ are time varying. 

We now introduce the triggering function. Denote the error at time $k$ between the updated state $x_i[k+1]$ and the auxiliary state $\hat{x}_i(k)$ by $e_i[k] = \hat{x}_i[k] - x_i[k+1]$ for $k \geq 0$. Then, let
\begin{equation}
f_i[k] = |e_i[k]| - (c_0 + c_1[k]  ), \label{trigger}
\end{equation}
where $c_0 \geq 0$ is a constant and $c_1[k]$ takes nonnegative and decreasing values with $c_1[k] \rightarrow 0$ in finite time. 
The roles of $c_0$ and $c_1[k]$
are to reduce the triggering frequency, and especially $c_1[k]$ allows the threshold to be large in the initial phase. Each node $i$ will
check this function and whenever it finds $f_i[k]$ to be positive, it will transmit its new state $x_i[k+1]$ to its neighbors.

We employ the control input taking account of possible delays in the transmission. Thus, we extend \eqref{input_syn} as
\begin{equation}
u_i[k]=\sum_{j\in \mathcal{N}_i^{l-}} a_{ij}[k](\hat{x}_j^P[k-\tau_{ij}^P[k]]-x_i[k]),  \label{input_asyn}
\end{equation}
where $\hat{x}_j^P[k]$ denotes the value of node $j$ at time $k$ sent along path $P$ and $\tau_{ij}^P[k]\in \mathbb{Z}_+$ denotes the delay in this $(j,i)$-path $P$ at time $k$.
The delays are time varying and may be different in each path. We assume the common upper bound $\tau$ on any \textit{normal} path $P$, over which all internal nodes are normal, as
\begin{equation}
0\leq \tau_{ij}^P[k] \leq \tau,\medspace j\in \mathcal{N}_i^{l-}, \medspace k\in \mathbb{Z}_+.
\end{equation}
In the following part, we also assume that every normal node $i$ updates its value at least once in every $\theta\geq 1$ steps. When $\theta= 1$, updates are synchronous.
%, where $\theta>\tau$ (This is not necessary but we assume this for improving the communication efficiency). 
%Hence, each normal node $i$ becomes aware of the updated value (no matter the value is triggered or not) of each of its normal $l$-hop neighbor $j$ on each normal $(j,i)$-path $P$ at least once in $\theta$ time steps, but possibly at different time instants. 
Although we impose this bound on the delays for message transmissions, the normal nodes need neither the value of this bound nor the information whether a path $P$ is a normal one or not. Also, there is no constraint on the size of $\tau$.

Under the delay bound $\tau$ imposed in \eqref{input_asyn}, triggered values of each node must reach all the multi-hop neighbors in $\tau$ steps. We have two possible relay models that can be employed in the proposed multi-hop algorithm:

%In the proof of Theorem \ref{theorem1}, we consider the delay bound $\tau$ for any multi-hop neighbors, if the node can send its triggered value to all the multi-hop neighbors in $\tau$ steps, then we consider such relay model is effective. (The value of $\tau$ can be relatively large.) Therefore, we have the two possible relay models as follows:

(i) \textit{Periodic relay model:} Each node relays all the recently received messages to its one-hop neighbors every $\lambda$ steps. If $\lambda=1$, each node must immediately relay the received messages. This is referred to as the \textit{immediate relay model}.

(ii) \textit{Package relay model:} Each node relays all the recently received messages along with its own values (e.g., in a message package) to its one-hop neighbors when its own event is triggered.

Among the two modes, clearly, the package relay model requires less frequent message transmissions and may be a more natural model in the event-based algorithm studied here.  We note however that with this model, it must be assumed that at time $k=0$, the neighboring agents exchage their state values. This is to cope with the situation where no event is triggered by any of the agents. This can occur since the event triggering function only takes account of the local states. We will illustrate the difference of the effects of the two relay models through simulations later.

%For the package relay model, if the relaying messages are sent at the event-triggering instants, then there is a possibility that the value update for every node is ``balanced'', and they do not send triggered values since no events have been triggered. This results in the dysfunction of the multi-hop relay model, then the case is simplified to the one-hop case and multi-hop communication is not utilized in this case. However, this issue can be easily solved if each node have the information on the initial transmitted values $\hat{x}_j[0]$ of every multi-hop neighbors. We will illustrate the difference of the effects of the two relay models through simulations.

\subsection{Threat Model}\label{threatmodel}

Next, we introduce the threat model studied here.
\begin{definition}
	\textit{($f$-total/$f$-local set)}
	The set of adversary nodes $\mathcal{A}$ is said to be $f$-total
	if it contains at most $f$ nodes, i.e., $\left| \mathcal{A}\right| \leq f$.
	Similarly, it is said to be $f$-local (in $l$-hop neighbors)
	if any normal node $i\in \mathcal{N}$ has at most $f$ adversary nodes as its $l$-hop neighbors, i.e., $\left|\mathcal{N}_i^{l-} \cap \mathcal{A}\right| \leq f, \forall i \in \mathcal{N}$.
\end{definition}

\begin{definition}
	\textit{(Byzantine nodes)}
	An adversary node $i\in \mathcal{A}$ is said to be a Byzantine node if it can arbitrarily modify its own value and relayed values, and moreover, it can send different values to its neighbors at each iteration.\footnote[2]{Here a Byzantine node can also decide not to send any value. This behavior corresponds to the omissive/crash model.} 
\end{definition}

As commonly done in the literature, 
we assume that each normal node knows the value of $f$ and the topology information of the graph up to $l$ hops. 
In the multi-hop setting, it is important to impose the following assumption.

\begin{assumption}
	%Each malicious node $i$ can manipulate its own state $x_i[k]$ and the values in the messages that they relay, but cannot change the path values in such messages.
	Each Byzantine node $i$ cannot manipulate the path values in the 
	messages containing its own state $x_i[k]$ and those that it relays. 
\end{assumption}

This is introduced for ease of analysis, but is not a strong constraint. In fact, manipulating message paths can
be easily detected and hence does not create problems. See the discussions in \cite{yuan2021resilient}.

\subsection{Resilient Asymptotic Consensus}\label{sec:resilient-asymptotic-consensus}

We now introduce the type of consensus among the normal agents to be sought in this paper.

\begin{definition}
	Given $c\geq 0$,
	if for any possible sets and behaviors of the
	adversary agents and any state values of the normal
	nodes, the following two conditions are satisfied,
	then we say that the normal agents reach 
	resilient consensus at the error level $c$:  %$\delta$
	
	\begin{enumerate}
		\item Safety: There exists a bounded safety interval $\mathcal{S}$ determined by the initial values of the normal agents such that $x_i[k] \in \mathcal{S}, \forall i \in \mathcal{N}, k \in \mathbb{Z}_+$. 
		%The set $\mathcal{S}$ is called the safety interval.
		\item Agreement: For all $i, j \in \mathcal{N}$, it holds that $\limsup_{k\to \infty}|x_i[k]-x_j[k]|\leq c$.
	\end{enumerate}
	
\end{definition}

\vspace{0.3cm}

\section{Event-triggered Algorithm Design}

In this section, we outline the structure of the event-triggered multi-hop weighted MSR (MW-MSR) algorithm. Then we define the strongly robust graphs with $l$ hops, which is crucial for guaranteeing Byzantine consensus \cite{yuan2022asynchronous}.

\subsection{Asynchronous Event-triggered MW-MSR algorithm}

At each time $k$, each normal node $i$ updates as follows:

\textit{1.~Receive step:} Node $i$ receives neighbors' values through different paths (described in \eqref{input_asyn}) and chooses to update its state or not. If it chooses to update, then it proceeds to step~2. Otherwise, it keeps its value as $x_i[k+1]=x_i[k]$.
%Node $i$ receives neighbors' values and chooses to update or not. If it chooses not to update, then it keeps its value as $x_i[k+1]=x_i[k]$. Otherwise, it obtains the values most recently received from neighbors through different paths (described in \eqref{input_asyn}) and its own value $x_i[k]$.

\textit{2.~Update step:} Node $i$ updates its value $x_i[k+1]$ according to Algorithm 1 using the values most recently received from neighbors and its own value $x_i[k]$.
%Node $i$ updates its value $x_i[k+1]$ according to the MW-MSR algorithm in Algorithm~1.

\textit{3.~Transmit step:} Node $i$ checks the value of $f_i[k]$ and sets the value of $\hat{x}_i[k+1]$ as
\begin{equation}
\hat{x}_i[k+1] = \left\{
\begin{aligned} 
&x_i[k+1], & &\textup{if} \ f_i[k]>0,\\
&\hat{x}_i[k],  & &\textup{otherwise}.   \label{trigger_update}
\end{aligned}
\right.
\end{equation}
Here, the auxiliary variable will be updated only when the current value has varied enough to exceed a threshold, and only at this time the node sends its value and the relayed values over each $l$-hop path to node $j \in\mathcal{N}_i^{l+}$.

In the Transmit step and Receive step, the nodes exchange messages with others that are up to $l$ hops away. Then in the Update step, node $i$ updates its state using Algorithm~1. Note that the adversary nodes may deviate from this specification as we describe in the next subsection.

%To further reduce the amount of each transmission when one event is triggered, the node can send the relayed values only for those has changed since last event.

One important feature here to further reduce the amount of data in each transmission when an event is triggered is to require that the nodes can send only the relayed values that have changed since last event.

\begin{algorithm}[t] %\footnotesize
	\caption{MW-MSR Algorithm} 
	
	1) At time $k$, normal node $i$ obtains the most recently received messages of the nodes in $\mathcal{N}_i^{l-}$ and itself, whose set is denoted by $\mathcal{M}_i[k]$, and sorts the values in $\mathcal{M}_i[k]$ in an increasing order.
	
	2) (a) Define two subsets of $\mathcal{M}_i[k]$ based on the message values:
	%	$\overline{\mathcal{M}}_i[k]=\{ m\in \mathcal{M}_i[k]: \mathrm{value}(m)> x_i[k-1]  \}$, 
	%	
	%	$\underline{\mathcal{M}}_i[k]=\{ m\in \mathcal{M}_i[k]: \mathrm{value}(m)< x_i[k-1]  \}$.
	\begin{equation*}
	\overline{\mathcal{M}}_i[k]=\{ m\in \mathcal{M}_i[k]: \mathrm{value}(m)> x_i[k]  \},
	\end{equation*}
	\begin{equation*}
	\underline{\mathcal{M}}_i[k]=\{ m\in \mathcal{M}_i[k]: \mathrm{value}(m)< x_i[k]  \}.
	\end{equation*}
	
	(b) Then, let $\overline{\mathcal{R}}_i[k]=\overline{\mathcal{M}}_i[k]$ if the cardinality of a minimum cover of $\overline{\mathcal{M}}_i[k]$ is less than $f$, i.e., $\left|  \mathcal{T}^* (\overline{\mathcal{M}}_i[k])\right| <f$. Otherwise, let $\overline{\mathcal{R}}_i[k]$ be the largest sized subset of $\overline{\mathcal{M}}_i[k]$ such that (i) for all $m\in \overline{\mathcal{M}}_i[k]\setminus \overline{\mathcal{R}}_i[k]$ and $m'\in \overline{\mathcal{R}}_i[k]$ we have $\mathrm{value}(m) \leq \mathrm{value}(m')$, and (ii) the cardinality of a minimum cover of $\overline{\mathcal{R}}_i[k]$ is exactly $f$, i.e., $\left|  \mathcal{T}^* (\overline{\mathcal{R}}_i[k])\right| =f$. 
	
	(c) Similarly, we can get $\underline{\mathcal{R}}_i[k]$ from $\underline{\mathcal{M}}_i[k]$, which contains the smallest values.
	
	(d) Finally, let $\mathcal{R}_i[k]=\overline{\mathcal{R}}_i[k]\cup\underline{\mathcal{R}}_i[k]$.
	%	\begin{equation*}
	%	\mathcal{R}_i[k]=\overline{\mathcal{R}}_i[k]\cup\underline{\mathcal{R}}_i[k].
	%	\end{equation*}

	3) Node $i$ updates its value as follows:
	\begin{equation}
	x_i[k+1]=\sum_{m\in \mathcal{D}_i[k]} a_{i}[k]\mathrm{value}(m),  \label{msrupdate}
	\end{equation}
	where $a_{i}[k]=1/\left| \mathcal{D}_i[k] \right| $ and $\mathcal{D}_i[k]=\mathcal{M}_i[k]\setminus \mathcal{R}_i[k]$.
	% and it satisfies $a_{ij}[k]\geq \alpha$, $0<\alpha<1$ is a constant.

\end{algorithm}

\subsection{The Notion of Strongly Robust Graphs}

The notion of graph robustness was first introduced in \cite{leblanc2013resilient}, and it was proved that graph robustness gives a tight condition guaranteeing resilient consensus using MSR-based algorithms. In \cite{yuan2021resilient}, we generalized this notion to the multi-hop case, where nodes can exchange values with their $l$-hop neighbors through different paths. Its definition is as follows.
%Note that for the malicious model, the necessary and sufficient condition to guarantee resilient consensus using MSR-based algorithms is that graph $\mathcal{G}$ is $(f+1,f+1)$-robust with $l$ hops \cite{leblanc2013resilient}, \cite{yuan2021resilient}.

\begin{definition}\label{rs-robust} A directed graph $\mathcal{G} = (\mathcal{V},\mathcal{E})$ is said to be $(r,s)$-robust with $l$ hops with respect to a given set $\mathcal{F}\subset \mathcal{V}$,
	if for every pair of nonempty disjoint subsets $\mathcal{V}_\text{1},\mathcal{V}_\text{2}\subset \mathcal{V}$, at least one of the following conditions holds:
	
	(1) $\mathcal{Z}_{\mathcal{V}_1}^r=\mathcal{V}_1$; 
	(2) $\mathcal{Z}_{\mathcal{V}_2}^r=\mathcal{V}_2$;
	(3) $\left| \mathcal{Z}_{\mathcal{V}_1}^r\right| +\left| \mathcal{Z}_{\mathcal{V}_2}^r\right| \geq s$,
	
	%	\begin{enumerate}
	%		\item $\mathcal{Z}_{\mathcal{V}_1}^r=\mathcal{V}_1$;
	%		\item $\mathcal{Z}_{\mathcal{V}_2}^r=\mathcal{V}_2$;
	%		\item  $\left| \mathcal{Z}_{\mathcal{V}_1}^r\right| +\left| \mathcal{Z}_{\mathcal{V}_2}^r\right| \geq s$;
	%	\end{enumerate}
	
	\noindent where $\mathcal{Z}_{\mathcal{V}_a}^\textit{r}$ is the set of nodes in $\mathcal{V}_\textit{a}$ ($a=1,2$) that have at least $r$ independent paths of at most $l$ hops originating from nodes outside $\mathcal{V}_\textit{a}$ and all these paths do not have any nodes in set $\mathcal{F}$ as intermediate nodes (i.e., the nodes in $\mathcal{F}$ can be source or destination nodes in these paths).
	Moreover, if the graph $\mathcal{G}$ satisfies this property with respect to any set $\mathcal{F}$ satisfying the $f$-total model, then we say that $\mathcal{G}$ is $(r,s)$-robust with $l$ hops (under the $f$-total model). 
\end{definition}

Intuitively speaking, for any set $\mathcal{F}\subset \mathcal{V}$ and for node $i\in \mathcal{V}_\text{1}$ to have the above-mentioned property, 
they should satisfy two conditions: (i) At least $r$ source nodes outside $\mathcal{V}_\text{1}$;
(ii) at least one independent path of length at most $l$ hops from each of the $r$ source nodes to node $i$, where such a path does not contain any internal nodes from the set $\mathcal{F}$.

%there should be at least $r$ source nodes outside $\mathcal{V}_\text{1}$ and at least one independent path of length at most $l$ hops from each of the $r$ source nodes to node $i$, where such a path does not contain any internal nodes from the set $\mathcal{F}$.

To deal with the Byzantine model, we need to focus on the subgraph consisting of only the normal nodes.
For the one-hop algorithms in \cite{leblanc2013resilient} and \cite{vaidya2012iterative}, the graph condition that the normal network is $(f+1)$-robust is proved to be necessary and sufficient for achieving resilient consensus under $f$-total Byzantine model.
In \cite{yuan2022asynchronous}, we extended this notion to the multi-hop setting and defined it as $r$-strongly robust graph with $l$ hops. Its definition is given as follows.

\begin{definition}\label{stronglyrobust}
	%\textit{($r$-strongly robustness with $l$ hops)}
	Let $\mathcal{F}$ be a subset of vertices in $\mathcal{G}$ and denote the subgraph of $\mathcal{G}$ induced
	by vertex set $\mathcal{V}\setminus\mathcal{F}$ as $\mathcal{G}_{\mathcal{V}\setminus\mathcal{F}}$.
	Then graph $\mathcal{G}$ is said to be $r$-strongly robust with $l$ hops with respect to $\mathcal{F}$ if the induced subgraph $\mathcal{G}_{\mathcal{V}\setminus\mathcal{F}}$ is $r$-robust with $l$ hops.
	If graph $\mathcal{G}$ satisfies this property with respect to any set $\mathcal{F}$ satisfying the $f$-total/local model, then we say that $\mathcal{G}$ is $r$-strongly robust with $l$ hops under the $f$-total/local model. When it is clear from the context, we just say $\mathcal{G}$ is $r$-strongly robust with $l$ hops.
	
\end{definition}

\begin{figure}[t]
	\centering
	\subfigure[]{
		\includegraphics[width=0.8in]{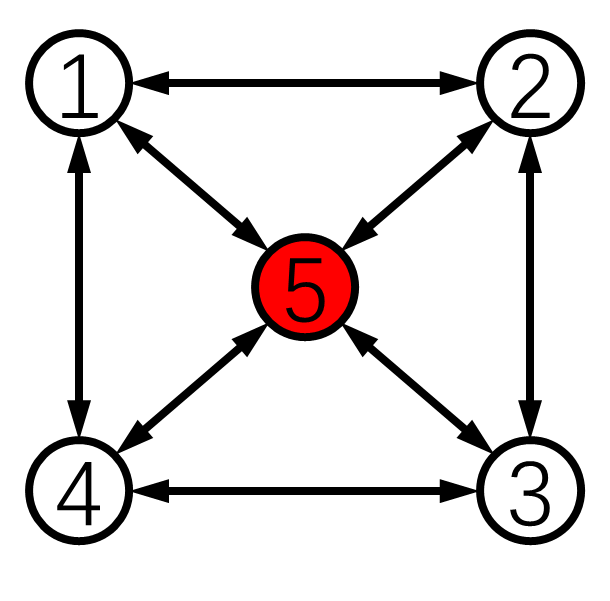}
		
	}
	\quad
	\vspace{-3pt}
	\subfigure[]{
		\includegraphics[width=1.2in]{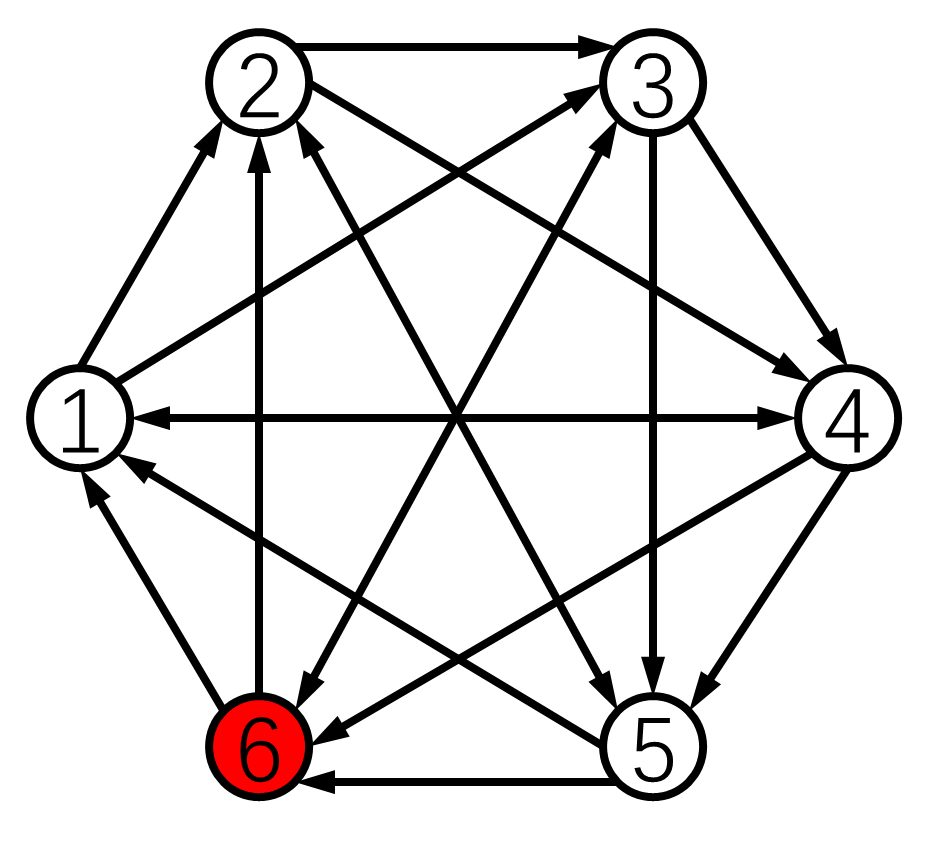}
		
	}
	\vspace{-3pt}
	\caption{ (a) The graph is not $2$-strongly robust with one hop but is $2$-strongly robust with $2$ hops. (b) The graph is $2$-strongly robust with one hop.}
	\label{1lcoal}
	\vspace*{-3.5mm}
\end{figure}

Generally, robustness of a graph increases as the relay range $l$ increases. See the examples in Fig.~\ref{1lcoal}.
Note that graph robustness with multi-hop communication needs to be checked for every possible set $\mathcal{F}$ satisfying the $f$-total/local model. 
We remark that the level of robustness is constrained by the in-degrees of the nodes. For instance, to achieve resilient consensus under the $f$-total malicious model, the minimum in-degree of the nodes needs to be at least $2f$. On the other hand, under the $f$-local Byzantine model, the minimum in-degree of the nodes is at least $2f+1$.

\section{Consensus Analysis}

In this section, we first prove the convergence of the asynchronous event-triggered MW-MSR algorithm. Then we discuss the effects of different relay models on the performance of the proposed algorithm.

To prove the convergence, 
we introduce two kinds of minimum and maximum of the states
of the normal agents. Denote the state vector and the transmitted state vector of normal agents at time $k$ by $x^N[k]$ and $\hat{x}^N[k]$, respectively. 

First, we denote the minimum and maximum of the states of the normal agents from time $k-\tau$ to time $k$ as
\begin{equation}
\begin{array}{lll} 
\overline{x}_\tau[k] =\max \left( x^N[k], x^N[k-1],\dots, x^N[k-\tau]\right),\\  \label{xtau}
\underline{x}_\tau[k] = \min \left( x^N[k], x^N[k-1],\dots, x^N[k-\tau]\right),
\end{array}
\end{equation}
respectively.
Next, we denote the joint minimum and maximum of the states and the transmitted states of the normal agents from time $k-\tau$ to time $k$, respectively, as
\begin{equation}
\begin{array}{lll} 
\overline{\hat{x}}_\tau[k] =\max \left( x^N[k], \dots, x^N[k-\tau], \hat{x}^N[k], \dots, \hat{x}^N[k-\tau]\right),\\
\underline{\hat{x}}_\tau[k] = \min \left( x^N[k], \dots, x^N[k-\tau], \hat{x}^N[k], \dots, \hat{x}^N[k-\tau]\right).
\end{array}
\end{equation}
%The safety interval $\mathcal{S}$ is chosen as $\mathcal{S} = [\underline{\hat{x}}_\tau[0],\overline{\hat{x}}_\tau[0]]$. 

We are ready to state the main theorem of the paper.

\begin{theorem}\label{theorem1}
	Consider a directed graph $\mathcal{G} = (\mathcal{V},\mathcal{E})$ with $l$-hop communication, where each normal node updates its value according to the asynchronous event-triggered MW-MSR algorithm.
	Under the $f$-local Byzantine model, the normal nodes reach resilient consensus at an
	error level $c$ if and only if the underlying graph is $(f+1)$-strongly robust with $l$ hops. Moreover, the safety interval is given by $\mathcal{S} = [\underline{\hat{x}}_\tau[0],\overline{\hat{x}}_\tau[0]]$, and the consensus error level $c$ is achieved if the parameter $c_0$ in the triggering function \eqref{trigger} satisfies
	\begin{equation}
	c_0\leq \frac{\gamma^{N\theta}}{4N\theta }c .
	\end{equation}
\end{theorem}

\vspace{0.3cm}

%\begin{proof}
    \textit{Proof:}
	\textit{(Necessity)} This part follows from our previous work \cite{yuan2022asynchronous}, which considers the special case without the triggering function, that is, $c_0 = c_1[k] = 0$.
	
	\textit{(Sufficiency)} 
	First, we show by induction that the safety condition is satisfied. 
	Note that the update rule \eqref{msrupdate} in Algorithm~1 can be rewritten as
	\begin{equation}
	x_i[k+1]=a_{i}[k]x_i[k]+\sum_{j\in \mathcal{D}_i[k]} a_{i}[k]\hat{x}_j^P[k-\tau_{ij}^P[k]],  \label{update_asyn}
	\end{equation}
	where $a_{i}[k]=1/\left| \mathcal{D}_i[k] \right| $. At time $k = 0$, it is clear
	by definition that $x_i[0], \hat{x}_i[0]\in \mathcal{S}$. 
	We first show that $\overline{\hat{x}}_\tau[k]$ is nonincreasing in time.
	From \eqref{update_asyn}, we have $x_i[k+1]\leq \overline{\hat{x}}_\tau[k]$ for all $i\in \mathcal{N}$ since the values larger than $\overline{\hat{x}}_\tau[k]$ are ignored in step 2 of Algorithm~1. Moreover, by \eqref{trigger_update}, it
	follows that $\hat{x}_i[k+1]\leq \overline{\hat{x}}_\tau[k]$ for all $i\in \mathcal{N}$. Together, we have $\overline{\hat{x}}_\tau[k+1]\leq \overline{\hat{x}}_\tau[k]$. We can similarly prove that $\underline{\hat{x}}_\tau[k]$ is nondecreasing in time. 
	%Thus, we have shown the safety condition.

	%Suppose that for each regular agent $i$, $x_i[k], \hat{x}_j^P[k-\tau_{ij}^P[k]] \in \mathcal{S}$. Then, for agent $i$, its neighbors' values in $\mathcal{M}_i[k]\setminus \mathcal{R}_i[k]$ must be in $\mathcal{S}$, since there are agents with values outside $\mathcal{S}$ at most $f$, and they are ignored in step 2 of Algorithm~1. From \eqref{update_asyn}, we have $x_i[k+1]\in \mathcal{S}$. Moreover, by \eqref{trigger_update}, it follows that $\hat{x}_i[k+1]\in \mathcal{S}$. Thus, the safety condition is satisfied. 
	
	We next show the consensus part. Note that for time $k\in (t_h^j,t_{h+1}^j)$ between two triggering instants, we have
	$f_i[k]\leq 0$. Moreover, for the neighbor node $j\in \mathcal{N}_i^{l-}$, if $f_j[k]> 0$, then we have $\hat{x}_j[k+1] = x_j[k+1]$. If $f_j[k]\leq 0$, then $\hat{x}_j[k+1] = \hat{x}_j[k] = x_j[k+1] + e_j[k]$. As a result, it holds
	$\hat{x}_j[k]= x_j[k] + \hat{e}_j[k-1]$ for $k \geq 1$, where
	\begin{equation}
	\hat{e}_j[k] = \left\{
	\begin{aligned} 
	&e_j[k], & &\textup{if} \ f_i[k]\leq 0,\\
	&0,  & &\textup{otherwise}.  
	\end{aligned}
	\right.
	\end{equation}
	Note that
	\begin{equation}
	|e_j[k]| \leq c_0 + c_1[k], \ \forall k\geq 0. \label{upper_e}
	\end{equation}
	Then, we can write \eqref{update_asyn} as
	\begin{equation}
	\begin{aligned} 
	&x_i[k+1]=   a_{i}[k]x_i[k] \\[2mm]
	& +\sum_{j\in \mathcal{M}_i[k]\setminus \mathcal{R}_i[k]} a_{i}[k](x_j^P[k-\tau_{ij}^P[k]]+\hat{e}_j[k-\tau_{ij}^P[k]-1]).   \label{xkplus1}
	\end{aligned}
	\end{equation}
	This can be bounded as 
	\begin{equation}
	\begin{aligned} 
	&x_i[k+1] \leq  a_{i}[k]\overline{x}_\tau[k] \\[3mm]
	&\hspace{0.5cm} +\sum_{j\in \mathcal{M}_i[k]\setminus \mathcal{R}_i[k]}  a_{i}[k](\overline{x}_\tau[k]+\hat{e}_j[k-\tau_{ij}^P[k]-1])\\[1mm]
	&\medspace \leq  \overline{x}_\tau[k] + \max_{j\in \mathcal{M}_i[k]\setminus \mathcal{R}_i[k]} |\hat{e}_j[k-\tau_{ij}^P[k]-1]|.
	\end{aligned}
	\end{equation}
	Thus, by \eqref{upper_e}, letting $c_1[k]=c_1[0]$ for $k<0$, we have 
	\begin{equation}
	x_i[k+1] \leq \overline{x}_\tau[k]+ c_0 + c_1[k-\tau-1].  \label{upper}
	\end{equation}
	Similarly, we have 
	\begin{equation}
	x_i[k+1] \geq \underline{x}_\tau[k]- c_0 - c_1[k-\tau-1].
	\end{equation}
	
	Let $V[k]= \overline{x}_\tau[k] - \underline{x}_\tau[k]$. Then, define two sequences by
	\begin{equation}
	\begin{aligned} 
	\overline{x}_0[k+1]&=  \overline{x}_0[k] + c_0 + c_1[k-\tau-1], \\
	\underline{x}_0[k+1]&=  \underline{x}_0[k] - c_0 - c_1[k-\tau-1],    \label{triger_convergence}
	\end{aligned}
	\end{equation}
	where $\overline{x}_0[0] = \overline{x}_\tau[0] -\sigma_0 $, and $\underline{x}_0[0] = \underline{x}_\tau[0] +\sigma_0$ with $\sigma_0 =\sigma V[0]$. 
	Then the following inequalities hold:
	\begin{equation}
	\begin{aligned} 
	\overline{x}_\tau[k]&\leq  \overline{x}_0[k] + \sigma_0 , \\  \label{tauzero}
	\underline{x}_\tau[k]&\geq  \underline{x}_0[k] - \sigma_0.   
	\end{aligned}
	\end{equation}
	We show $\overline{x}_\tau[k]\leq  \overline{x}_0[k] + \sigma_0$ by induction, and $\underline{x}_\tau[k]\geq  \underline{x}_0[k] - \sigma_0$ can be proved in a similar way. When $k=0$, we clearly have $\overline{x}_\tau[0]=  \overline{x}_0[0] + \sigma_0$. Suppose that \eqref{tauzero} holds. Then, we have at time $k + 1$
	\begin{equation}
	\begin{aligned} 
	\overline{x}_\tau[k+1] &=\max \left( x^N[k+1], x^N[k],\dots, x^N[k+1-\tau]\right)\\
	&\leq  \overline{x}_\tau[k] + c_0 + c_1[k-\tau-1] \\
	&\leq  \left( \overline{x}_0[k] + \sigma_0 \right) + c_0 + c_1[k-\tau-1]  \\
	&=  \overline{x}_0[k+1] + \sigma_0.
	\end{aligned}
	\end{equation}
	The first inequality holds because from \eqref{upper}, we have $\max x^N[k+1]\leq  \overline{x}_\tau[k] + c_0 + c_1[k-\tau-1]$. 
	Moreover, from \eqref{xtau}, we have $\max \left( x^N[k],\dots, x^N[k+1-\tau]\right) \leq  \overline{x}_\tau[k] $.

	We next introduce another sequence $\varepsilon_0[k]$ defined by
	\begin{equation}
	\varepsilon_0[k+1]= \gamma\varepsilon_0[k]-(1-\gamma)\sigma_0,  \label{epsilon}
	\end{equation}
	where $\varepsilon_0[0] = \varepsilon V[0]$. Take the positive $\varepsilon$ and $\sigma$ so that
	\begin{equation}
	\varepsilon+\sigma=\frac{1}{2},    \   0<\sigma<\frac{\gamma^{N\theta}}{1-\gamma^{N\theta}} \varepsilon.  \label{sigma}
	\end{equation}
	Here, we claim that it holds 
	\begin{equation}
	0 < \varepsilon_0[k+1] < \varepsilon_0[k],    \   k=0,1,\dots, N\theta-1.
	\end{equation}
	This is proved as follows. Since $0<\gamma<1$, from \eqref{epsilon}, we can easily have $\varepsilon_0[k+1] < \varepsilon_0[k]$. 
	It is thus sufficient to show $\varepsilon_0[N\theta]>0$. From \eqref{epsilon}, we have
	\begin{equation*}
	\begin{aligned} 
	\varepsilon_0[N\theta]&= \gamma^{N\theta}\varepsilon_0[0]-\sum_{j=0}^{N\theta-1} \gamma^j(1-\gamma)\sigma_0  \\
	&=  \left( \gamma^{N\theta}\varepsilon-(1-\gamma^{N\theta})\sigma \right) V[0].    
	\end{aligned}
	\end{equation*}
	This is positive because we have chosen $\varepsilon$ and $\sigma$ as in \eqref{sigma}.

	For the sequence $\varepsilon_0[k]$, define two sets as
	\begin{equation*}
	\begin{aligned}
	\mathcal{Z}_1(k,\varepsilon_0[k])&=\{i\in \mathcal{N}: x_i[k]>\overline{x}_0[k]-\varepsilon_0[k]\},\\
	\mathcal{Z}_2(k,\varepsilon_0[k])&=\{i\in \mathcal{N}: x_i[k]<\underline{x}_0[k]+\varepsilon_0[k]\}.
	\end{aligned}
	\end{equation*}
	These sets are both nonempty at time $k = 0$ and, in particular,
	each contains at least one normal node; this is because, by
	definition, $\overline{x}_\tau[0] > \overline{x}_0[0]-\varepsilon_0[0]$ and $\overline{x}_\tau[0] < \underline{x}_0[0]+\varepsilon_0[0]$.
	
	In the following, we show that $\mathcal{Z}_1(k,\varepsilon_0[k])$ and $\mathcal{Z}_2(k,\varepsilon_0[k])$ are disjoint sets. To this end, we must show
	\begin{equation}
	\overline{x}_0[k]-\varepsilon_0[k]\geq \underline{x}_0[0]+\varepsilon_0[0].  \label{disjoint}
	\end{equation}
	By \eqref{triger_convergence} for $\overline{x}_0[k]$ and $\underline{x}_0[k]$, we have
	\begin{equation*}
	\begin{aligned} 
	&(\overline{x}_0[k]-\varepsilon_0[k]) - (\underline{x}_0[k]+\varepsilon_0[k]) \\[2mm]
	&=\left( \overline{x}_0[0] +c_0k+\sum_{j=-\tau-1}^{k-\tau-2}c_1[j] \right) \\
	&\medspace\medspace - \left( \underline{x}_0[0] -c_0k-\sum_{j=-\tau-1}^{k-\tau-2}c_1[j] \right) - 2\varepsilon_0[k].
	\end{aligned}
	\end{equation*}
	Since $\overline{x}_0[0] = \overline{x}_\tau[0] -\sigma_0 $ and $\underline{x}_0[0] = \underline{x}_\tau[0] +\sigma_0$ with $\sigma_0 =\sigma V[0]$, we have 
	\begin{equation*}
	\begin{aligned} 
	&(\overline{x}_0[k]-\varepsilon_0[k]) - (\underline{x}_0[k]+\varepsilon_0[k]) \\[2mm]
	&=\left( \overline{x}_\tau[0] - \underline{x}_\tau[0]\right) - 2\sigma_0 + 2c_0k + 2\sum_{j=-\tau-1}^{k-\tau-2}c_1[j] - 2\varepsilon_0[k] \\
	&=V[0]-2\sigma V[0] + 2c_0k + 2\sum_{j=-\tau-1}^{k-\tau-2}c_1[j] - 2\varepsilon_0[k] \\
	&> (1-2\sigma-2\varepsilon)V[0] + 2c_0k + 2\sum_{j=-\tau-1}^{k-\tau-2}c_1[j] \geq 0.
	\end{aligned}
	\end{equation*}
	The last inequality holds since $\varepsilon+\sigma=1/2$ and $\varepsilon_0[k] < \varepsilon_0[0] = \varepsilon V[0]$. Thus, we have proved  \eqref{disjoint}.

	So far, we have shown that the two sets $\mathcal{Z}_1(k,\varepsilon_0[k])$ and $\mathcal{Z}_2(k,\varepsilon_0[k])$ are disjoint. 
	Notice that the network is $(f + 1)$-strongly robust with $l$ hops w.r.t. any set $\mathcal{F}$ following the $f$-local model and the set of Byzantine nodes $\mathcal{A}$ also satisfies the $f$-local model. Hence, the network is $(f + 1)$-strongly robust with $l$ hops w.r.t. the set $\mathcal{A}$ and at least one of the conditions in Definition \ref{rs-robust} for robustness holds.
	Therefore, if the two sets are both nonempty, then for these two nonempty disjoint sets $\mathcal{Z}_1(k,\varepsilon_0[k])$ and $\mathcal{Z}_2(k,\varepsilon_0[k])$, one of them has a normal agent with at least $f +1$ independent normal paths originating from some normal nodes outside.
	
	Suppose that normal node $i\in \mathcal{Z}_1(k,\varepsilon_0[k])$ has the above-mentioned property. A similar argument holds when $i\in \mathcal{Z}_2(k,\varepsilon_0[k])$.
	Now, we go back to the update rule \eqref{xkplus1} for node $i$ and rewrite it by partitioning the neighbor set 
	into two parts: those that belong to $\mathcal{Z}_1(k,\varepsilon_0[k])$ and those that do not. Node $i$ has at least $f +1$ independent normal paths originating from the normal nodes outside. According to Algorithm~1, it will use at least one value originating from the normal nodes outside $\mathcal{Z}_1(k,\varepsilon_0[k])$; thus, we obtain 
	\begin{equation*}
	\begin{aligned} 
	&x_i[k+1]  =   a_{i}[k]x_i[k]  +\sum_{j\in \mathcal{D}_i[k]\cap\mathcal{Z}_1} a_{i}[k]x_j^P[k-\tau_{ij}^P[k]]  +\\[2mm]
	&\sum_{j\in \mathcal{D}_i[k]\setminus \mathcal{Z}_1} a_{i}[k]x_j^P[k-\tau_{ij}^P[k]] +\sum_{j\in \mathcal{D}_i[k]} a_{i}[k]\hat{e}_j[k-\tau_{ij}^P[k]-1]\\
	&\leq a_{i}[k]\overline{x}_\tau[k]  +\sum_{j\in \mathcal{D}_i[k]\cap\mathcal{Z}_1} a_{i}[k]\overline{x}_\tau[k] + \\[2mm]
	&\sum_{j\in \mathcal{D}_i[k]\setminus \mathcal{Z}_1} a_{i}[k](\overline{x}_0[k]-\varepsilon_0[k]) +\sum_{j\in \mathcal{D}_i[k]} a_{i}[k]\hat{e}_j[k-\tau_{ij}^P[k]-1].\\
	\end{aligned}
	\end{equation*}
	Combining \eqref{tauzero} and the fact that $a_{i}[k]$ is lower bounded by $\gamma$, we have 
	\begin{equation}
	\begin{aligned} 
	&x_i[k+1]  \leq (1-\gamma)\overline{x}_\tau[k]  + \gamma(\overline{x}_0[k] - \varepsilon_0[k]) + c_0 + c_1[k-\tau-1] \\[1mm]
	&\leq (1-\gamma)(\overline{x}_0[k] + \sigma_0)  + \gamma(\overline{x}_0[k] - \varepsilon_0[k]) + c_0 + c_1[k-\tau-1] \\[1mm]
	&\leq \overline{x}_0[k] + c_0 + c_1[k-\tau-1] + (1-\gamma)\sigma_0 - \gamma\varepsilon_0[k] \\[1mm]
	&= \overline{x}_0[k+1] - \varepsilon_0[k+1]    \label{convergence}
	\end{aligned}
	\end{equation}
	for $k=0,1,\dots, N\theta-1$, where the first inequality follows from the assumption that $\mathcal{Z}_1(k,\varepsilon_0[k])$ is nonempty, and the equality follows from \eqref{triger_convergence} and \eqref{epsilon}. The relation in \eqref{convergence} shows that once an update happens at node $i$, then this node will move out of $\mathcal{Z}_1(k+1,\varepsilon_0[k+1])$. It is further noted that inequality \eqref{convergence} also holds for the normal nodes that are not in $\mathcal{Z}_1(k,\varepsilon_0[k])$ at time $k$. This indicates that the nodes outside $\mathcal{Z}_1(k,\varepsilon_0[k])$ will not move in $\mathcal{Z}_1(k+1,\varepsilon_0[k+1])$. Similar results hold for the set $\mathcal{Z}_2(k+1,\varepsilon_0[k+1])$.
	
	Recall that the normal nodes update at least once for every $\theta$ steps. 
	As a result, if the two sets $\mathcal{Z}_1(k,\varepsilon_0[k])$ and $\mathcal{Z}_2(k,\varepsilon_0[k])$ are both nonempty at time $k$, then after $N\theta$ time steps, all the normal nodes will be out of at least one of them. Suppose that $\mathcal{Z}_1(k,\varepsilon_0[k])$ is empty. When such an event occurs at $k=0$, it clearly follows that $\overline{x}_\tau[N\theta] \leq \overline{x}_0[N\theta] - \varepsilon_0[N\theta]$. From the definition of $V[k]$, we have
	\begin{equation*}
	\begin{aligned} 
	V&[N\theta]  =\overline{x}_\tau[N\theta]-\underline{x}_\tau[N\theta] \\[1mm]
	&\leq \left( \overline{x}_0[N\theta] - \varepsilon_0[N\theta] \right)  - \left( \underline{x}_0[N\theta] - \sigma_0 \right) \\[1mm]
	&=\overline{x}_0[0] - \underline{x}_0[0] + 2c_0N\theta + 2\sum_{j=-\tau-1}^{N\theta-\tau-2}c_1[j] - \varepsilon_0[N\theta] + \sigma_0 \\
	&=  \left(\overline{x}_\tau[0] - \sigma_0 \right) - \left(\underline{x}_\tau[0] + \sigma_0 \right)  + 2c_0N\theta + 2\sum_{j=-\tau-1}^{N\theta-\tau-2}c_1[j]\\
	&\medspace \medspace \medspace - \varepsilon_0[N\theta] + \sigma_0 \\
	&=V[0]-\sigma V[0] + 2c_0N\theta + 2\sum_{j=-\tau-1}^{N\theta-\tau-2}c_1[j]  \\
	&\medspace \medspace \medspace - \left( \gamma^{N\theta}\varepsilon- (1-\gamma^{N\theta})\sigma \right) V[0]\\
	&= \left( 1-\gamma^{N\theta}(\varepsilon+\sigma) \right) V[0] + 2c_0N\theta + 2\sum_{j=-\tau-1}^{N\theta-\tau-2}c_1[j].
	\end{aligned}
	\end{equation*}
	
	By \eqref{sigma}, we have
	\begin{equation}
	V[N\theta] \leq \left( 1-\frac{\gamma^{N\theta}}{2} \right) V[0] + 2c_0N\theta + 2\sum_{j=-\tau-1}^{N\theta-\tau-2}c_1[j].  \label{vnt}
	\end{equation}
	If there are more updates by node $i$ after time $k = N\theta$, this argument can be extended further as
	\begin{equation}
		\begin{aligned} 
			&V[hN\theta]  \leq \left( 1-\frac{\gamma^{N\theta}}{2} \right) V[(h-1)N\theta]  \\
			&\hspace{1.5cm}  +  2c_0N\theta + 2\sum_{j=(h-1)N\theta-\tau-1}^{hN\theta-\tau-2}c_1[j].\\
		\end{aligned}
	\end{equation}
    Hence, we have
	\begin{equation}
	\begin{aligned} 
	&V[hN\theta] \leq \left( 1-\frac{\gamma^{N\theta}}{2} \right)^h V[0]  + \sum_{t=0}^{h-1} \left( 1-\frac{\gamma^{N\theta}}{2}\right)^{h-1-t} \\
	&\hspace{1.5cm} \times  \left(2c_0N\theta + 2\sum_{j=tN\theta-\tau-1}^{(t+1)N\theta-\tau-2}c_1[j]\right)\\
	&\leq \left( 1-\frac{\gamma^{N\theta}}{2} \right)^h V[0]  + 2c_0N\theta \frac{1-\left(1-\frac{\gamma^{N\theta}}{2}\right)^h}{1-\left(1-\frac{\gamma^{N\theta}}{2}\right)}\\
	&\medspace \medspace \medspace +  \sum_{t=0}^{h-1} \left( 1-\frac{\gamma^{N\theta}}{2}\right)^{h-1-t} \left( 2\sum_{j=tN\theta-\tau-1}^{(t+1)N\theta-\tau-2}c_1[j]\right).   \label{vhnt}
	\end{aligned}
	\end{equation}
	Since $c_1[k] \rightarrow 0$ in finite time, there exists a finite time $h_0$ such that $c_1[k] = 0, k \geq h_0N\theta$. Then, for $h\geq h_0$, we can obtain from \eqref{vhnt}
	\begin{equation}
	\limsup_{h\to \infty} V[hN\theta]  \leq \frac{2c_0N\theta}{1-\left(1-\frac{\gamma^{N\theta}}{2}\right)} = \frac{4c_0N\theta}{\gamma^{N\theta}}\leq c.
	\end{equation}
	The analysis is similar for the dynamics of $V[hN\theta+t]$, $t =0, 1,\dots,N\theta - 1$, and we obtain as in \eqref{vhnt}:
	\begin{equation*}
	\limsup_{h\to \infty} V[hN\theta+t]  \leq  \frac{4c_0N\theta}{\gamma^{N\theta}}\leq c. \tag*{$\blacksquare$}
	\end{equation*}
%\hfill  $\blacksquare$
%\end{proof}

\begin{figure}[t]
	\centering
	%\vspace{-10pt}
	\subfigure[\scriptsize{One-hop case without delays.}]{
		\includegraphics[width=3.4in,height=1.6in]{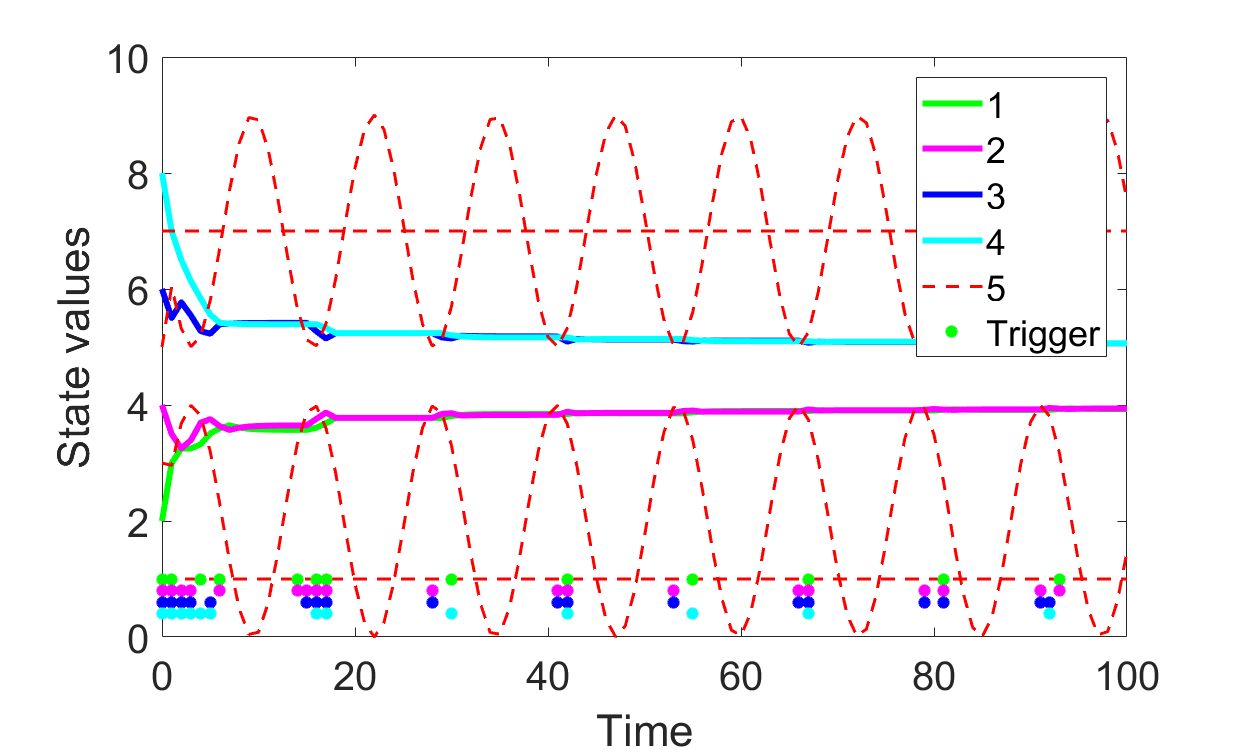}
	}
	
	\vspace{-10pt}
	\subfigure[\scriptsize{Two-hop case with delays.}]{
		\includegraphics[width=3.4in,height=1.6in]{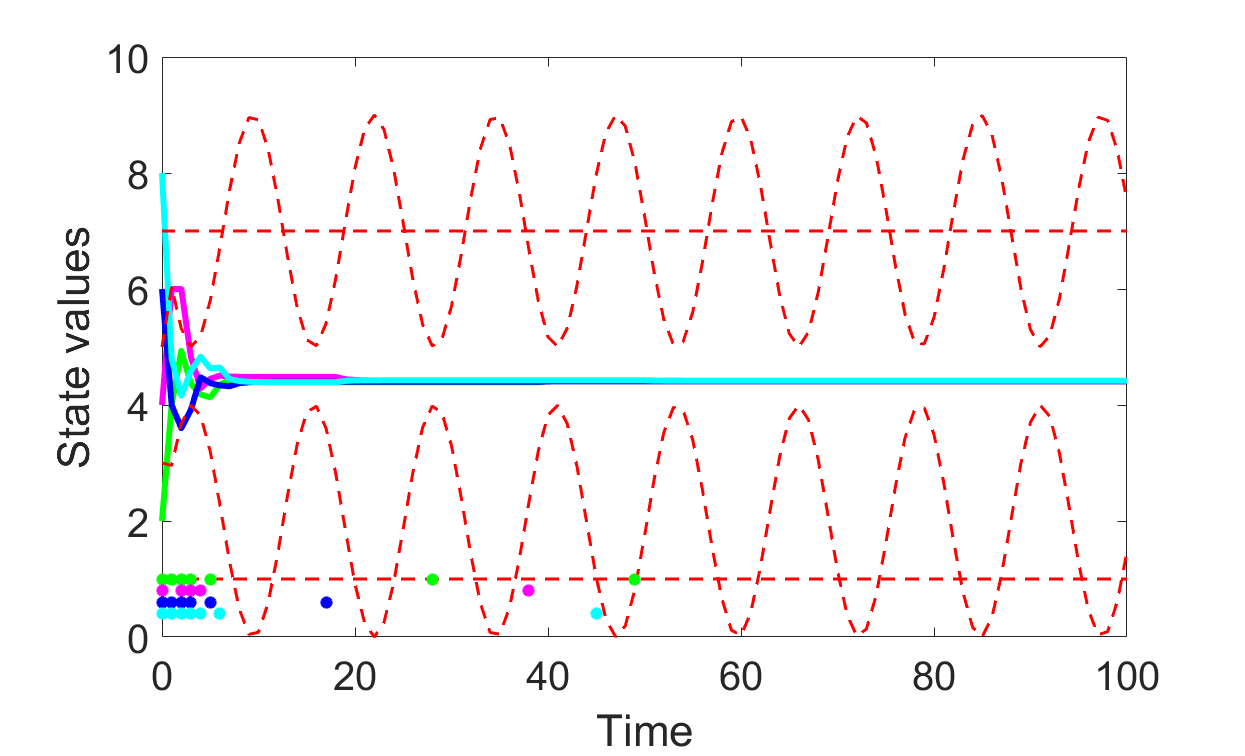}
	}
	
	\caption{Time responses using different event-triggered MSR algorithms.}
	\label{response_5}
	\vspace*{-3.5mm}
\end{figure}

As we can see from \eqref{vnt}, the delays make the consensus error bigger than the one under no delays for every $N\theta$ steps, i.e., the term containing $c_1[k]$ is bigger that the one under no delays. However, when the iteration number is large enough as in \eqref{vhnt}, the term containing $c_1[k]$ converges to 0, which results in the same error bound $c$ as the one under no delays in the one-hop case \cite{wang2019resilient}. This fact shows that although delays can slow down the consensus process, they do not affect the consensus error bound as also observed in \cite{dibaji2017resilient}, \cite{yuan2021resilient}.

\section{Numerical Examples}

\begin{figure}[t]
	\centering
	
	%\vspace{-10pt}
	%\captionsetup{font={scriptsize}}
	\subfigure[\scriptsize{One-hop case without delays.}]{
		\includegraphics[width=3.4in,height=1.6in]{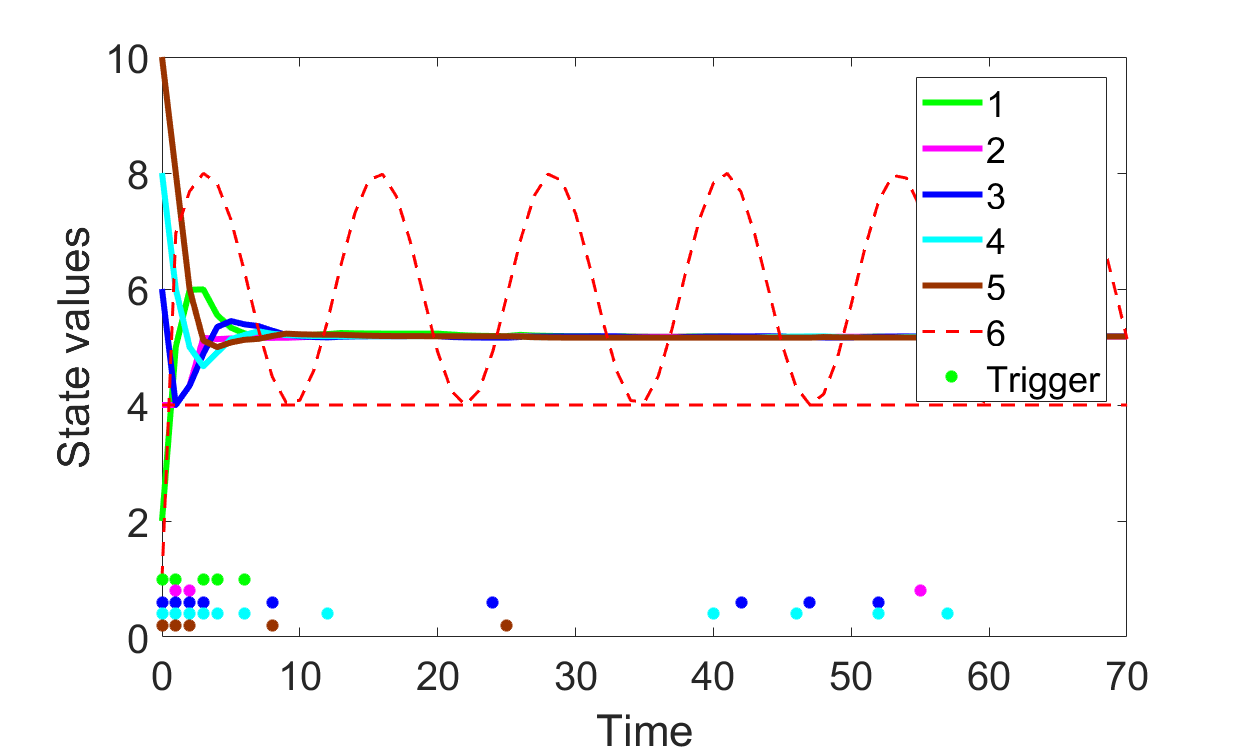}
	}

	\vspace{-10pt}
	\subfigure[\scriptsize{Two-hop case with immediate relays.}]{
		\includegraphics[width=3.4in,height=1.6in]{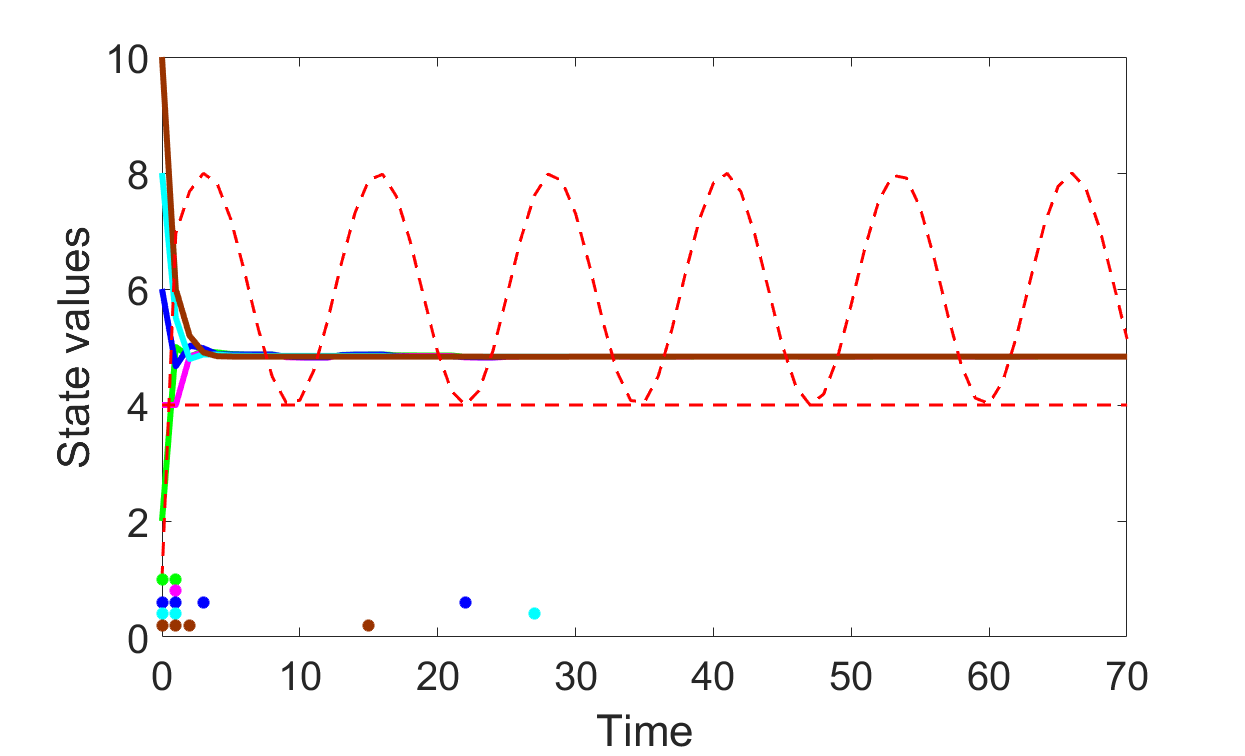}
	}

	\vspace{-10pt}
	\subfigure[\scriptsize{Two-hop case with package relays.}]{
		\includegraphics[width=3.4in,height=1.6in]{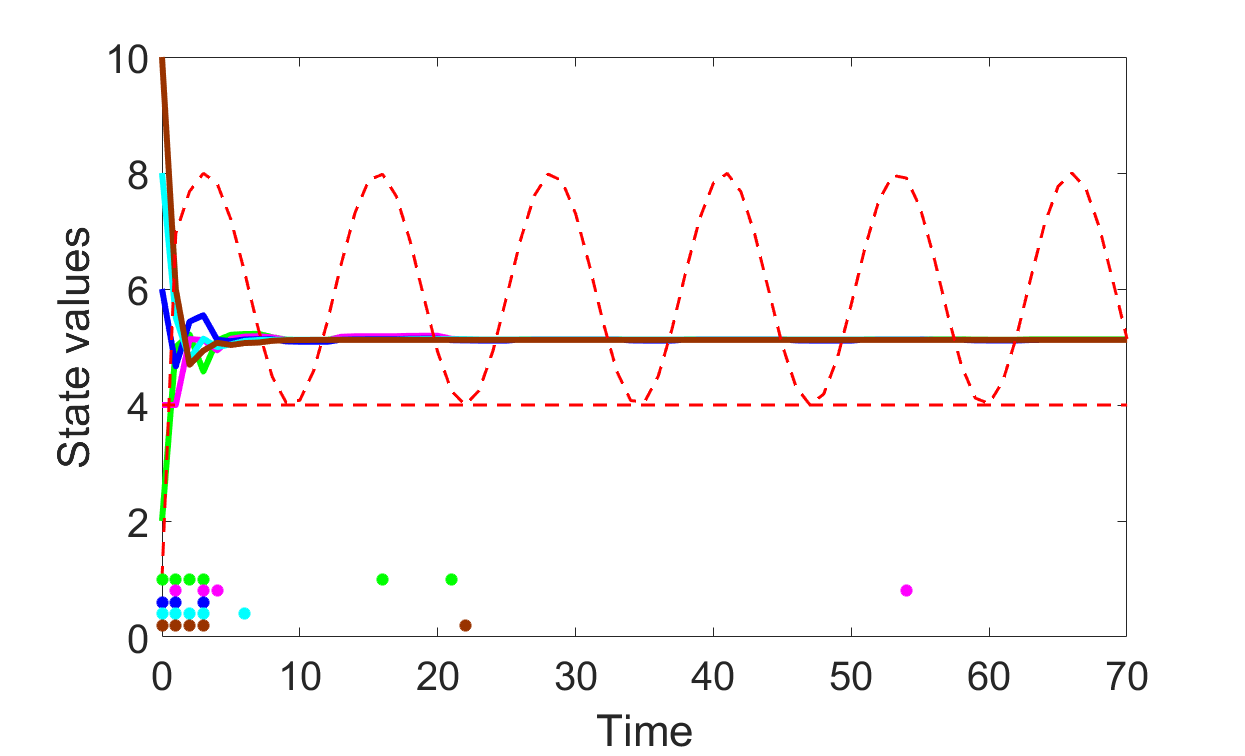}
	}
	%\vspace{-7pt}
	\caption{Time responses using different event-triggered MSR algorithms.}
	\label{response_6}
	\vspace*{-3.5mm}
\end{figure}

In this section, we conduct simulations for networks applying the event-triggered MW-MSR algorithm. For all the simulations, we set the parameters $c_0$ and $c_1[k]$ of the triggering function as $c_0=1.215\times10^{-2}$ and $c_1[k]=0.5\times e^{-0.06(k+20)}$, respectively.

\subsection{Topology Gap between One-hop and Multi-hop Algorithms}

In this part, we show that the proposed algorithm can guarantee resilient consensus in a network where the conventional one-hop algorithm cannot.
Consider the network in Fig.~\ref{1lcoal}(a). This graph is not $2$-strongly robust with one hop, but is with $2$ hops. Suppose that node 5 is Byzantine and sends four different values to its four neighbors. Let the initial normal states be $x^N[0]=[2\ 4\ 6\ 8]^T$. 
According to \cite{leblanc2013resilient}, \cite{wang2019resilient}, this graph does not meet the condition for $1$-total Byzantine model even for synchronous updates. Thus, resilient consensus is impossible as shown in Fig. \ref{response_5}(a) where the four red dashed lines indicate the adversarial values and the dots represent the time instants when events are triggered by the normal nodes.
%Then, we examine the synchronous two-hop MW-MSR algorithm under the same attack, and resilient consensus is achieved as shown in Fig. \ref{two-hop-syn}.

Then, we perform simulations for the asynchronous two-hop event-triggered MW-MSR algorithm under the same attacks. 
Let the normal nodes update synchronously with delays in communication ($\theta=1$). 
Moreover, we choose the package relay model, i.e., nodes only relay the messages when events are triggered at the nodes.
Observe that resilient consensus is achieved as shown in Fig. \ref{response_5}(b). This verifies the effectiveness of the proposed algorithm.
%Note that the safety interval $\Delta x_{\tau}[k]=\max z^N[k]- \min z^N[k]$ is nonincreasing. 

\subsection{The Amount of Transmissions of Different Algorithms}

\begin{table}[t]\label{table1}\scriptsize
	
	\caption{average triggering times per normal node}
	\renewcommand\arraystretch{1.8}
	\centering
	\vspace{-0.15cm}
	\begin{tabular}{ccc} 
		\toprule
		
		Algorithms &  Average events &  Average transmissions\\
		\midrule
		
		One-hop  & 7.26 & 7.26\\
		Two-hop with immediate relays & 3.05 & 12.20\\
		Two-hop with package relays & 6.99 & 6.99\\
		
		\bottomrule
		
	\end{tabular}
	\vspace*{-5mm}
\end{table}

In this part, we show that the amount of transmissions of the proposed algorithm can be further reduced compared to the one-hop algorithm.
This time, we consider the network in Fig.~\ref{1lcoal}(b). This graph is $2$-strongly robust with one hop, and hence, with $2$ hops (see \cite{yuan2021resilient}).
Node 6 is Byzantine and is capable to send two different values to its neighbors (including different relayed values). Let the initial normal states be $x^N[0]=[2\ 4\ 6\ 8\ 10]^T$. 
By \cite{leblanc2013resilient} and Theorem \ref{theorem1}, this graph satisfies the condition for $1$-total Byzantine model. Thus, resilient consensus can be achieved with both one-hop and two-hop algorithms, and the results are given in Fig. \ref{response_6}.
%Then, we examine the synchronous two-hop MW-MSR algorithm under the same attack, and resilient consensus is achieved as shown in Fig. \ref{two-hop-syn}.

From Fig. \ref{response_6}, we can also see that the numbers of events of the two-hop algorithm with immediate relays and package relays are both smaller than that of the one-hop algorithm. This is because by introducing the multi-hop communication, each node can have more information of the network, which may result in faster speed of the consensus process and less events. Moreover, observe that the two-hop algorithm with immediate relays has less events than the algorithm with package relays. Obviously, the immediate relay model is an ideal model and it requires additional communication resources for the relaying process. Note that for this model, each event is accompanied with additional transmissions for relays as each node has three neighbors. In contrast, the package relay model is more realistic and energy-saving since it requires only communication for the events, but reaching consensus takes longer. 

To verify these properties of the algorithms, we further conducted Monte Carlo simulations in the same network for 50 runs by randomly taking initial normal states within $[0,10]$. The Byzantine node 6 misbehaves as in the previous simulation. Table I displays the average times of events and transmissions per normal node of the three algorithms. In all runs, consensus was achieved and the results are consistent with our analysis so far. 
In particular, the package relay model requires the least number of transmissions overall.

\section{Conclusion}

In this paper, we have investigated the resilient consensus problem using the event-triggered MSR algorithm with multi-hop communication.
We have characterized the network requirement for the proposed algorithm to guarantee resilient consensus with a certain error level.
We found that the delays in communication may slow down the consensus process, but they do not affect the consensus error.
By introducing multi-hop communication, even sparse graphs can meet the condition for robustness. Furthermore, the event-triggered scheme provides an effective way to reduce the number of transmissions for the multi-hop communication.

\addtolength{\textheight}{-12cm}

\end{document}